\documentclass[pdflatex,sn-mathphys-num]{sn-jnl}

\usepackage{graphicx}%
\usepackage{multirow}%
\usepackage{amsmath,amssymb,amsfonts}%
\usepackage{amsthm}%
\usepackage{mathrsfs}%
\usepackage[title]{appendix}%
\usepackage{xcolor}%
\usepackage{textcomp}%
\usepackage{manyfoot}%
\usepackage{booktabs}%
\usepackage{algorithm}%
\usepackage{algorithmicx}%
\usepackage{algpseudocode}%
\usepackage{listings}%
\usepackage{graphicx}
\usepackage{float}

\theoremstyle{thmstyleone}%
\theoremstyle{thmstyletwo}%

\theoremstyle{thmstylethree}%

\begin{document}

\title[Article Title]{Enhancing Diagnostic Precision in Gastric Bleeding through Automated Lesion Segmentation: A Deep DuS-KFCM Approach}


\author[1]{\fnm{Xian-Xian} \sur{LIU}}\email{yc37972@um.edu.mo}

\author[2]{\fnm{Mingkun} \sur{XU}}\email{ xumingkun@gdiist.cn}

\author[3,4]{\fnm{Yuanyuan} \sur{WEI}}\email{wei-yy18@link.cuhk.edu.hk }

\author[5]{\fnm{Huafeng} \sur{QIN}}\email{qinhuafengfeng@163.com}

\author[6]{\fnm{Qun} \sur{SONG}}\email{ songqun@ctbu.edu.cn}

\author[1]{\fnm{Simon} \sur{FONG}}\email{ ccfong@umac.mo}

\author[7]{\fnm{Feng} \sur{TIEN}}\email{tianfeng@hebeu.edu.cn}

\author[8]{\fnm{Wei} \sur{LUO}}\email{luowei\_421@163.com}

\author[9]{\fnm{Juntao} \sur{GAO}}\email{jtgao@tsinghua.edu.cn }

\author[10]{\fnm{Zhihua} \sur{ZHANG}}\email{zhangzhihua@big.ac.cn}

\author[11]{\fnm{Shirley} \sur{SIU}}\email{ shirleysiu@mpu.edu.mo}

\affil*[1]{\orgdiv{The Department of Computer and Information Science}, \orgname{University of Macau}, 
\orgaddress{\city{Macau SAR}, 
\postcode{999078}, \country{ China}}}

\affil*[2]{\orgname{Guangdong Institute of Intelligence Science and Technology}, \orgaddress{\city{Zhuhai}, \postcode{519031}, \country{China}}}

\affil[3]{\orgdiv{Department of Biomedical Engineering}, \orgname{The Chinese University of Hong Kong}, \orgaddress{\street{Shatin}, 
\postcode{999077}, \state{Hong Kong SAR}, \country{China}}}

\affil[4]{\orgdiv{Department of Neurology, David Geffen School of Medicine}, \orgname{University of California}, \orgaddress{\city{Los Angeles},
\postcode{90095}, \state{California}, \country{USA}}}

\affil[5]{\orgdiv{The School of Computer Science and Information Engineering}, \orgname{Chongqing Technology and Business University}, \orgaddress{\city{Chongqing},
\postcode{400067}, \country{China}}}

\affil[6]{\orgdiv{Institute of Artificial Intelligence}, \orgname{Chongqing Technology and Business University}, \orgaddress{\city{Chongqing},
\postcode{400067}, \country{China}}}

\affil[7]{\orgdiv{Hebei Key Laboratory of Medical Data Science, Institute of Biomedical Informatics, School of Medicine}, \orgname{Hebei University of Engineering}, \orgaddress{\city{Handan, Hebei Province},
\postcode{056038}, \country{China}}}

\affil[8]{\orgdiv{The director of the Institute of Clinical Medicine}, \orgname{The First People's Hospital of Foshan}, \orgaddress{\city{Guangzhou},
\postcode{510060}, \country{China}}}

\affil[9]{\orgdiv{The Beijing National Research Center for Information Science and Technology (BNRist) }, \orgname{Tsinghua University}, \orgaddress{\city{Beijing},
\postcode{100084}, \state{California}, \country{China}}}

\affil[10]{\orgdiv{zhangzhihua@big.ac.cn}, \orgname{Beijing Institute of Genomics, Chinese Academy of Sciences}, \orgaddress{\city{Beijing},
\postcode{100101}, \country{China}}}

\affil[11]{\orgdiv{Centre for Artificial Intelligence Driven Drug Discovery}, \orgname{Macao Polytechnic University, Rua de Luís Gonzaga Gomes}, \orgaddress{\city{Macau},
\postcode{999078}, \country{China}}}


\abstract{Timely and precise classification and segmentation of gastric bleeding in endoscopic imagery are pivotal for the rapid diagnosis and intervention of gastric complications, which is critical in life-saving medical procedures. Traditional methods grapple with the challenge posed by the indistinguishable intensity values of bleeding tissues adjacent to other gastric structures. Our study seeks to revolutionize this domain by introducing a novel deep learning model, the Dual Spatial Kernelized Constrained Fuzzy C-Means (Deep DuS-KFCM) clustering algorithm. This Hybrid Neuro-Fuzzy system synergizes Neural Networks with Fuzzy Logic to offer a highly precise and efficient identification of bleeding regions. Implementing a two-fold coarse-to-fine strategy for segmentation, this model initially employs the Spatial Kernelized Fuzzy C-Means (SKFCM) algorithm enhanced with spatial intensity profiles and subsequently harnesses the state-of-the-art DeepLabv3+ with ResNet50 architecture to refine the segmentation output. Through extensive experiments across mainstream gastric bleeding and red spots datasets, our Deep DuS-KFCM model demonstrated unprecedented accuracy rates of 87.95\%, coupled with a specificity of 96.33\%, outperforming contemporary segmentation methods. The findings underscore the model's robustness against noise and its outstanding segmentation capabilities, particularly for identifying subtle bleeding symptoms, thereby presenting a significant leap forward in medical image processing.}

\keywords{ Fuzzy clustering (FC) segmentation, Quantified eigenvalue, lesion detection, Hybrid Neuro-Fuzzy, GLCM }



\maketitle

\section{Introduction}\label{sec1}

The accurate detection of lesions is paramount in the early diagnosis and subsequent patient management of gastric cancer (EGC), a perilous condition prevalent across the globe. The stakes are high in the domain of gastric health; the stomach is an organ where detection delays can drastically reduce survival odds. While early interventions can offer survival rates soaring above 96\% \cite{cite1}, \cite{cite2}, \cite{cite3}, \cite{cite4}, \cite{cite5}, late-stage detections grimly plummet to a mere 20-40\% five-year survival rate  \cite{cite6}. Thus, refining diagnostic precision to segregate healthy individuals from those at risk becomes a mission of critical importance. 

Analyzing and processing endoscopy gastric bleeding lesion areas are the most challenging and upcoming field. The complexity of segmenting bleeding tissues is further compounded by the similarity in intensity values between pathological and adjacent healthy tissues, coupled with the challenge of distinguishing features amid image noise. Image segmentation is entailed with the division or separation of the image into regions of similar features \cite{cite7}, \cite{cite8}, \cite{cite9}, \cite{cite10}, \cite{cite11}, \cite{cite12}, \cite{cite13}, \cite{cite14}. Traditional manual segmentation of endoscopy images is a very time-consuming task and subject to intra and inter-rater variability. Therefore, numerous researchers have worked and developed methods for solving cancer problems by using medical image segmentation. 

Recent advancements in automated image processing techniques have begun to bridge the gap in lesion detection capabilities. Among these, convolutional neural networks (CNN) have shown promise due to their ability to learn and illustrate complex features. For example, the multi-scale attention-guided convolutional neural network is presented by Zheng et al. (2022) to effectively capture the varied properties of gastric cancer lesions. The authors demonstrated improved segmentation accuracy compared to traditional methods, highlighting the potential of advanced machine learning techniques for this task\cite{cite15},  \cite{cite16},  \cite{cite17}, \cite{cite18} and Li et al. (2021) proposed a hybrid segmentation framework combining multiple approaches, including thresholding, region growing, and clustering, for the detection and delineation of gastric ulcers in endoscopic images. Their method addressed the challenges of varied lesion characteristics and showed promising results for clinical applications, yet the specificity for bleeding tissues in gastric images remains under-explored\cite{cite19}. 

Building on this literature, we have implemented a solution for segmenting images from the public dataset (available from https://datasets.simula.no/kvasir-seg/), which includes the gastric bleeding (GB) and gastric red spots (GRS) images. We have engineered a computer-aided model employing the innovative Deep DuS-KFCM clustering method. This hybrid approach harnesses deep learning's spatial acumen, merging it with an advanced clustering algorithm, to accurately delineate GB lesions and equip clinicians with both quantitative and qualitative assessments of their patients. Ultimately, 
our comparative analysis positions the Deep DuS-KFCM method as a frontrunner in this domain, demonstrating its superiority in segmentation accuracy and computational efficiency against contemporary benchmarks. The necessity of adopting deep learning techniques in clinical practice cannot be overstated, as they not only streamline the diagnostic process but also enhance the precision of lesion detection, ultimately leading to better patient management and improved clinical outcomes.

\section{Proposed methodology}\label{sec2}

Accurate segmentation of gastric lesions is essential for effective diagnosis and treatment planning. However, manual segmentation methods on the endoscopic images which are usually screen captures from a real-time video recording, face significant limitations due to their reliance on subjective clinician judgment, which can lead to substantial variability and inconsistencies in quantified assessment. The complexity of differentiating between gastric lesions and surrounding healthy tissues is compounded by similarities in pixel intensity, noise, and artifacts in endoscopic images. These challenges render manual segmentation not only labor-intensive but also susceptible to error, potentially delaying critical interventions and negatively impacting patient outcomes.

While recent advancements in automated image processing, particularly those utilizing deep learning approaches, have shown promise in enhancing segmentation accuracy, many existing methods still struggle with the intricacies of bleeding lesion detection. Traditional deep learning frameworks often focus on isolated features such as texture or color, limiting their effectiveness in distinguishing subtle variations between pathological and non-pathological tissues. This inadequacy highlights the need for a more comprehensive segmentation strategy that can incorporate both pixel-wise information and spatial context.
The proposed dual spatially kernelized constrained fuzzy c-means (DuS-KFCM) methodology is designed to overcome these limitations by integrating advanced fuzzy clustering techniques with a deep learning framework. By utilizing both pixel values and their spatial inter-relationships, DuS-KFCM enhances the algorithm's ability to identify and segment lesions accurately, addressing the shortcomings of conventional methods. This approach draws upon texture analysis through Gray Level Co-occurrence Matrix (GLCM) statistics, enriching the feature representation and enabling more robust differentiation of lesions from surrounding tissues.

Moreover, the DuS-KFCM framework not only improves segmentation precision but also enhances computational efficiency, positioning it as a valuable tool for clinical practice. By automating the segmentation process and reducing the reliance on subjective assessments, this methodology aims to facilitate quicker and more accurate diagnoses of gastric conditions.
As we turn to the specifics of the DuS-KFCM methodology, it is important to emphasize how this innovative approach integrates fuzzy clustering algorithms with deep learning techniques. This synergy offers a solution specifically designed to address the complex challenges associated with gastric bleeding lesions. The framework of the Deep DuS-KFCM method demonstrates its potential for improved segmentation accuracy and computational efficiency compared to existing methods.

\subsection{Stage \#1:  Introduction of Dus-KFCM}\label{subsec2}

In stage 1, this research introduces a novel, composite method fuzzy DuS-KFCM, which relies on both pixel values and spatial inter-relationships for pixel-wise classification and segmentation of endoscopy images. Integrating the color and texture is more effective than solely relying on texture features, utilizing statistics of GLCM and SKFCM clustering to represent images \cite{cite20}, \cite{cite21} , \cite{cite22}. The image processing method presented in this section, DuS-KFCM, allows a fuzzy region image segmentation, drawing inspiration from SKFCM but incorporating GLCM texture analysis induced by compression from normal and abnormal (Kernelized Constrained Deep Fuzzy Clustering based on Dual Spatial Information)\cite{cite23}. 

\begin{figure}[htbp]
\centering
\includegraphics[width=0.9\textwidth]{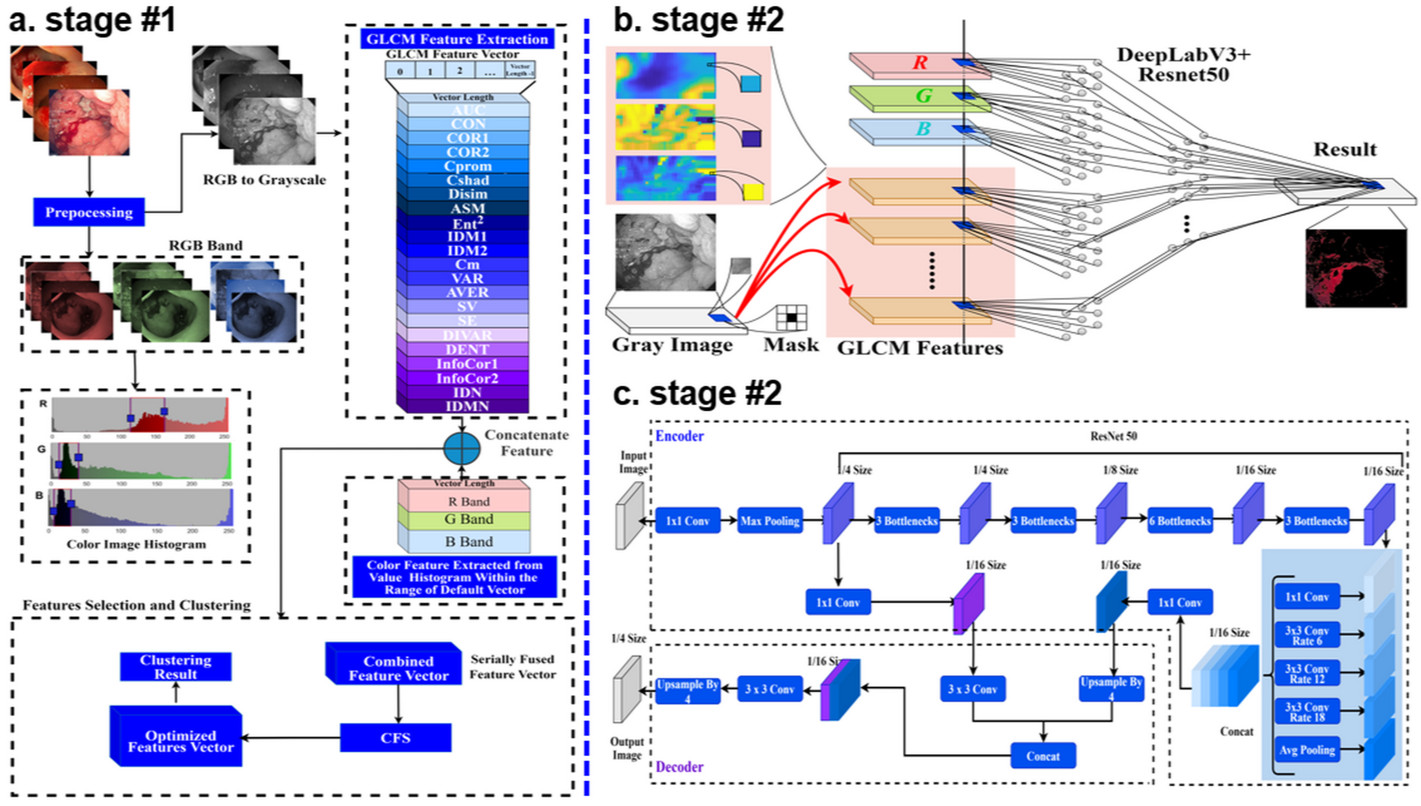}
\caption{\textbf{Model Overview of Deep DuS-KFCM: (a) Overview of the Feature Extraction System using DuS-KFCM. }This figure illustrates the initial phase of segmentation with the DuS-KFCM method, highlighting the algorithm's process for distinguishing potential gastric lesions from surrounding tissues by analyzing GLCM and fuzzy clustering for GB feature classification. The workflow encapsulates the progress from lesion identification to cancer boundary specification, aiming for efficient classification.\textbf{ (b) }\textbf{Network Architecture Diagram: DeepLabv3+ with ResNet50 for Enhanced Gastric Lesion Segmentation.} Showcasing the systemic architecture employed for refining the boundaries of segmented gastric lesions, this figure details how the DeepLabv3+ model, backed by ResNet50, navigates through the complexities of image segmentation-mitigating misclassification through precise lesion boundary refinement.\textbf{ (c) Pre-Segmentation and Refinement Process for Gastric Endoscopy Imaging with DLv3+ and ResNet50, Based on DuS-KFCM.} Providing a comprehensive view of the neural network's block diagram, this figure illustrates the dual-stage segmentation process. Initially leveraging the DuS-KFCM technique for basic lesion identification, it further demonstrates the deep learning model's region-enhancing capabilities, highlighting the encoder and decoder modules' roles in achieving a refined and highly accurate lesion segmentation outcome. }\label{fig1}
\end{figure}

We initiate the segmentation process employing DuS-KFCM to initially identify potential lesions within the gastric images. This process benefits from the algorithm's ability to effectively differentiate lesions from surrounding gastric tissue despite the challenging similarity in pixel intensity values. The proposed method is used to find GLCM and Fuzzy Clustering for GB Features classification. The goal of the proposed method is to output useful information for cancer boundary through medical image segmentation and be efficient in classifying cancer \cite{cite24}, \cite{cite25}. This study also attempts to combine several methods to create an effective segmentation. A flowchart of this process is illustrated in Figure~\ref{fig1}(a).

The GLCM features are utilized to calculate 22 statistical features from the grayscale gastric image, encompassing parameters such as energy, contrast, entropy, correlation, and homogeneity. Additionally, the (R, G, B) color features are extracted from each visible band (R, G, B) histogram within the default vector range, acquired through training the lesion color feature. These computed features are then sequentially fused, selecting the most discriminative features optimized by CFS, resulting in the coarse clustering outcome. The primary objective behind developing this feature descriptor is to effectively capture both color and texture information present in the image. This feature serves as a versatile descriptor capable of accommodating a wide range of images from diverse databases \cite{cite26}, \cite{cite27}, \cite{cite28}, \cite{cite29}, \cite{cite30}, \cite{cite31}.

\subsection{Stage \#2: Implementation of the deep DuS-KFCM image segmentation}\label{subsec2}
Following this pre-segmentation, we apply a based on deep learning model to refine the gastric bleeding lesion segmentation boundaries in stage \#2. The proposal of a novel algorithm above in which a combination of color, texture, and shape features is used to form the feature vector. The feature vector that combines all features is used to cluster images into different clusters. The optimized ensemble enhancement of DuS-KFCM for medical image segmentation involves calculating the extracted GB-trained image region. Subsequently, high-level features are extracted from the segmented image to enhance the segmentation process \cite{cite32}, \cite{cite33}, \cite{cite34}, \cite{cite35}, \cite{cite36}. This ensemble approach captures a balance between high-level details and the broader, contextual image information critical to accurate lesion identification.

In our work, DeepLabv3+ with ResNet50 as the backbone is used as the segmentation network \cite{cite37},  \cite{cite38},  \cite{cite39}, \cite{cite40},  \cite{cite41},  \cite{cite42}, \cite{cite43}. A significant proportion of the misclassification rate may be attributed to the small spurious regions that are frequently included in the cluster corresponding to the lesion region. To overcome the aforementioned limitations, the deep DuS-KFCM method is implemented to improve the performance of the image segmentation process. DeepLabv3 is an effective decoder that helps to refine the segmentation of lesion boundaries. Figure~\ref{fig1}(b) displays the structure of the proposed system.

We start by taking each image from the training set and segmenting the gastric bleeding lesions from the image using the DuS-KFCM fuzzy clustering technique. We repeat the above process for every image in the data set and use them to train our model – deep DuS-KFCM model. Then, for every testing image, we segment the GB lesion using the same method and then classify the lesion as benign or malignant using our model. The aim of this research is to get the contours of interest approximation in the medical images determined by employing Dus-SKFCM and enhancement by deep learning model(region-growing) \cite{cite44}.

A block diagram representation of a neural network is shown in Figure~\ref{fig1}(c). Incorporating networks into information to refine the lesion segmentation results. The encoder module utilizes a pre-trained ResNet-50 network to extract features from the input image, capturing hierarchical representations at multiple scales. Atrous convolutions and the ASPP module enhance multi-scale contextual information. The decoder module refines segmentation by fusing high-level gastric lesion information with detailed spatial features, improving accuracy, especially at object boundaries. The integration of global context and spatial details improves localization and precision for comprehensive segmentation results.

\section{Criterion performance metrics}\label{sec3}

We employ criterion performance metrics, including accuracy, sensitivity, precision, Matthews correlation coefficient, Dice coefficient, Jaccard index, specificity, and intersection over union, to benchmark the quality of our clustering efforts. High segmentation accuracy is essential not only for quantified reporting but also for reducing inter-observer variability, thereby enhancing the reliability of our results.

The quality of the clustering can be assessed in two ways: directly and indirectly. In the direct method, we apply various cluster validity measures to evaluate whether the clustering quality is improving. In the indirect method, we utilize the proposed clustering approach to form data clusters, which are then employed to build classifiers. The performance of these classifiers serves as an indirect measure of clustering quality; however, this is only feasible when the data are labeled.

We assess clustering quality both directly, through cluster validity measures, and indirectly by leveraging these clusters to construct classifiers. This dual approach provides a comprehensive means to evaluate clustering efficacy.

Segmentation accuracy (SA) is defined as the ratio of the total number of correctly classified pixels to the total number of pixels. To quantitatively compare segmentation performance, it is computed as follows:
\begin{equation}
SA = \sum_{i=1}^{k} \frac{S_i \cap G_i}{n}
\end{equation}

where the S\textsubscript{i} represents a segmentation result, the G\textsubscript{i} denotes the corresponding ground truth, the k is the number of clusters and n is the total number of pixels of images.
Sensitivity is the amount of positive items correctly identified:
\begin{equation}
\text{Sens.} = \frac{TP}{TN + FN}
\end{equation}
where TP is the number of true positive, FP is the number of false positives and FN is the number of false negatives.

The precision is defined as the sum of TPs is divided by the sum of TPs and FPs. The performance of the precision is computed for segmented object, and it is capable of a returning only related instances to the classification model. The precision is expressed in (3):
\begin{equation}
\text{Precision} = \frac{TP}{TP + FP}
\end{equation}

F1-score is the function of precision and recall. It is calculated when a balance between precision and recall is needed.

\begin{equation}
F1 = 2 \times \frac{Precision \times Recall}{Precision + Recall}
\end{equation}

The Matthews correlation coefficient (MCC) is essentially a correlation coefficient between the actual classification and the predicted classification, which defined as follows:

\begin{equation}
MCC = \frac{TP_i \times TN_i - FP_i \times FN_i}{\sqrt{ (TP_i + FN_i)(TP_i + FP_i)(TN_i + FP_i)(TN_i + FN_i) }}
\end{equation}

Dice-Coefficien is a statistical measure of similarity rate between two sample sets:

\begin{equation}
\text{Dice}_{cof.} = \frac{2 \times TP}{2 \times TP + FP + FN}
\end{equation}

Jaccard Index is a measure of similarity rate between two sample sets: Jaccard similarity: GT represents ground truth, and CS denotes cancer segmentation, which defined as follows:

\begin{equation}
\text{Jac}_{idx} = \frac{TP}{TP + FP + FN}
\end{equation}

Specificity is the rate of correct identification of negative items. The specificity is expressed in (9).

\begin{equation}
\text{Spef.} = \frac{TN}{TN + FN}
\end{equation}

The intersection-over-union (IoU), also known as the Jaccard Index, is one of the most commonly used metrics in semantic segmentation.

\begin{equation}
\text{IoU} = \frac{X \cap Y}{X \cup Y}
\end{equation}

where X and Y are manually annotated segmentation mask and predicted segmentation mask, respectively.

\section{Method verification and practical applications }\label{sec4}

\begin{figure}[H]
\centering
\includegraphics[width=0.9\textwidth]{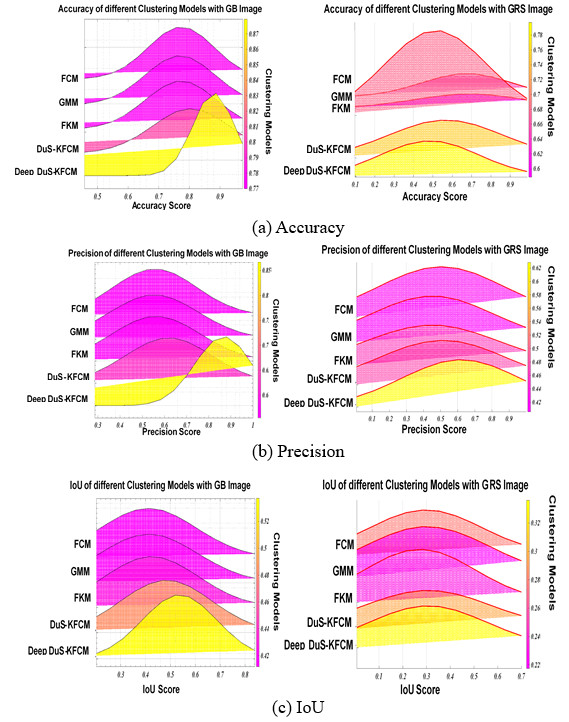}
\caption{\textbf{ }\textbf{Comparative Performance Analysis of Clustering Models on GB/GRS Database. }This figure portrays the reliability of our novel Deep DuS-KFCM model against traditional methods across both GB and GRS datasets. Key performance metrics—accuracy, precision, and IoU—under differential evaluation scenarios are plotted respectively.
}\label{fig2}
\end{figure}

\begin{figure}[htbp]
\centering
\includegraphics[width=0.9\textwidth]{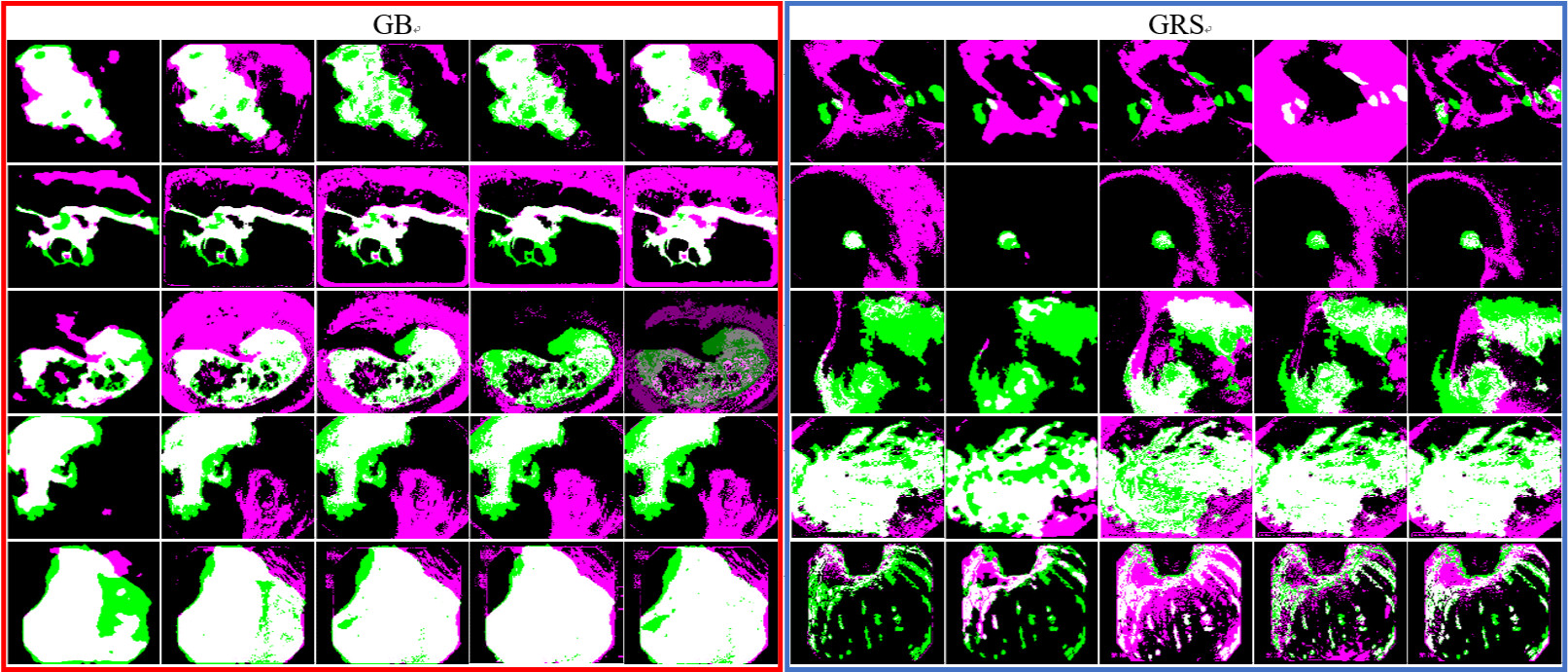}
\caption{\textbf{ }\textbf{\textbf{Segmentation Efficacy on Synthetic Pseudo-color Images}}\textbf{\textbf{.}} Displayed are the results for the GB and GRS datasets within red and blue delineations, respectively. Here, the juxtaposition of our deep DuS-KFCM method with traditional threshold techniques illuminates the finesse of its clustering accuracy. Lesion boundaries are discriminatively marked: system-detected margins are drawn in magenta, whereas gastroenterologist-annotated ground truths are green, and reference information-fused areas appear white. 
 }\label{fig3}
\end{figure}

Our discussions and results pivot around the superior segmentation capabilities demonstrated by our novel approach. Through strategic adaptation and testing, the deep DuS-KFCM model has set a new benchmark for the segmentation of early gastric bleeding cancer images, paving the way for advancements in computer-aided diagnostics. Benchmark tests and visual comparisons, particularly illustrated Figure~\ref{fig2}  and Figure~\ref{fig3} , which details the comparison of the proposed method and other prediction methods, like Fuzzy K-Means (FKM), Gaussian Mixture Model (GMM), and Fuzzy c-means (FCM), our innovative fuzzy set segmentation algorithm, on the gastric bleeding dataset and gastric red spot dataset, respectively, affirm our method’s enhanced capability. It can be noticed that the global accuracy, precision and IoU score results from the automatic segmentation’s performance evaluation, are all increased obviously after applying the preprocessing deep fuzzy enhancement step. The inclusion of a pre-processing step, dubbed 'deep fuzzy enhancement', markedly elevated our method's performance, this obvious quantitative enhancement assures the success of our proposed approach in batch automatic fuzzy segmentation. 

To provide a comprehensive overview of the quantitative metrics across GB images utilizing the DuS-KFCM models integrated into a novel fuzzy variational model for accurate lesion delineation, the detection and segmentation performance were evaluated using state-of-the-art comparison metrics. In terms of performance analysis of each object segmented image, to provide a detailed study, we have retrieved some more early gastric bleeding cancer images which is reported with the help of accuracy, precision, IoU and etc. in the blue box of Figure~\ref{fig3} . For each input image, our method shows better robustness and identification to the lesion area which maintain the equal or nearly equal high recognition rate for each. The segmentation visual comparison results for each GB image are illustrated in the left sub-figure of Figure~\ref{fig3}. Experimental results have been shown visually and achieve reasonable consistency. We achieved a remarkable accuracy of 98\% in endoscopy gastric images, solidifying our method's reliability. 

To demonstrate the effectiveness of the proposed method deep DuS-KFCM, we compared with other recently reported prediction methods on the similar gastric datasets. All the methods are performed using GRS cross-validation test. Our analysis extended to GRS images sourced from varied databases, by utilizing Gastric Red-Spots (GRS) images for cross-validation, our method's classification accuracy stood at 87.9475\%, surpassing previously explored methodologies. With visual Gaussian ridge on GRS images in the blue box of Figure~\ref{fig3}  showcasing a side-by-side comparison of segmented results against ground truths, the adeptness of our model in distinguishing lesions amid complex backgrounds was particularly noteworthy, as evidenced by the distinct segmentation areas color-coded in our visual presentations. The empirical results underscore the proposed method's superior performance, evidenced by consistently high accuracy and precision coefficients indicating its robustness in diverse imaging conditions. The experimental results demonstrate its utility in the classification of various gastric tissue features.

In summary, our proposed Deep DuS-KFCM method affirms its potential as a highly accurate, efficient solution for the segmentation of gastric bleeding lesions. Its deployment in a computer-aided diagnosis environment promises significant improvements in the clinical management of gastric conditions.

\section{Conclusion and perspectives}\label{sec5}

In our study, we have introduced a novel method for automated detection and segmentation of lesions in gastric bleeding images, employing a deep learning-enhanced fuzzy clustering approach-Deep DuS-KFCM. A comprehensive juxtaposition against manual annotations from the GB database conveys the proposed model's ability to extract lesions with an adeptness akin to expert analysis. This novel method outshines conventional cancer segmentation techniques, heralding a transformative stride forward for computational medical diagnostics.

From a clinical imaging standpoint, deep DuS-KFCM's utility is manifest across the varying types of endoscopic images encountered in gastric cancer diagnostics. In this method, three types of images such as endoscopy results for gastric bleeding cancer, normal, and gastric red-spots image. The proposed method achieved 87.95\% and 79.72\% for the accuracy coefficient whereas 86.69\% and 62.82\% for the precision coefficient, respectively. The accuracy and precision coefficients are utilized to compare the diversity and similarity pixels between the images. The model’s impressive accuracy and precision coefficients affirm its proficiency in cancer detection, further enhancing the computationally-aided diagnostic landscape. 

Looking to the future, our aspirations include refining the algorithm to detect and segment metastatic cancers of non-gastric origin and adapting its application beyond the medical field to genomic data clustering. However, notwithstanding the clinical impact, we grapple with consistent challenges such as the scarcity of meticulously labeled pathologic images, which underscores the significance of developing comprehensive datasets. Novelty in feature extraction and classifier design looms on the horizon as a burgeoning domain for investigation, propelling gastric cancer image analysis into new terrains of potential.

Our revelations point to a transformative future where the deep DuS-KFCM method not only etches the present landscape of precision in medical imaging but also ventures forth into the uncharted waters of clinical identification and taxonomy of gastric pathologies. Emphasizing diagnostic accuracy over computational expenditure, this trailblazing technique unfolds as a paradigmatic shift in cancer detection efficacy, symbiotically aligning with conventional clinical diagnostic processes.

\bmhead{Acknowledgements}

The authors would like to express their sincere gratitude to the Tsinghua Biology Information Directors of Laboratories for their valuable feedback and insightful suggestions throughout this research. We also extend our thanks to the reviewers for their constructive comments, which greatly enhanced the quality of this paper. Special appreciation goes to The First People's Hospital of Foshan, Guangzhou, for providing the sample gastric image dataset that was instrumental in the development and validation of our proposed method. Their support has been invaluable in advancing our research on automated detection and segmentation of gastric bleeding lesions.

\section*{Funding}

This work was supported by the Key-Area Research and Development Program of Guangdong Province (Grants No. 2021B0909060002), National Natural Science Foundation of China (Grants No.62204140), Guangzhou Development Zone Science and Technology (2021GH10, 2020GH10, 2023GH02), the University of Macau (MYRG2022-00271-FST), research grant by the Science and Technology Development Fund of Macau (0032/2022/A), Natural Science Foundation of Chongqing, China (Grant No. CSTB2022NSCQ-MSX1571) and the Science and Technology Research Program of Chongqing Municipal Education Commission (Grant No.KJQN202400841).

\section*{Data availability}

All data generated and analyzed during this study are included in the article and its supplementary information files.

\section*{Conflict of interest}

 The authors declare that they have no known competing financial interests or personal relationships that could have appeared to influence the work reported in this paper. 
 
\section*{Ethics approval}

This study that are considered exempt from ethical review are provided below: Studies that involve data available in the public domain.



\begin{thebibliography}{44}
\bibitem{cite1} Bertram, C. A. et al. Computer-assisted mitotic count using a deep learning-based algorithm improves interobserver reproducibility and accuracy. Vet. Pathol. 59, 211–226 (2022).
\bibitem{cite2}Huang, H.-Y. et al. Classification of skin cancer using novel hyperspectral imaging engineering via YOLOv5. \textit{J. Clin. Med.} 12, 1134 (2023). 
\bibitem{cite3}Siegel, R., Ma, J., Zou, Z. \& Jemal, A. Cancer statistics. \textit{CA Cancer J. Clin.} 64, 9–29 (2014).
\bibitem{cite4}Cancer statistics. \href{https://www.cancer.org/cancer/pancreatic-cancer/about/key-statistics.html}{https://www.cancer.org/cancer/pancreatic-cancer/about/key-statistics.html}, (2018). 
 
\bibitem{cite5} Sung, H. et al. Global cancer statistics 2020: Globocan estimates of incidence and mortality worldwide for 36 cancers in 185 countries. \textit{CA Cancer J. Clin.} 71, 209–249 (2021).

\bibitem{cite6}Misra, I. \& Maaten, L. V. D. Self-Supervised Learning of Pretext-Invariant Representations. \textit{Proceedings of the IEEE/CVF Conference on Computer Vision and Pattern Recognition (CVPR)}, 6707–6717 (2020).  

\bibitem{cite7}Lambin, P. et al. Radiomics: extracting more information from medical images using advanced feature analysis. \textit{Eur. J. Cancer} 48, 441–446 (2012). 
\bibitem{cite8}Eilaghi, A. et al. CT texture features are associated with overall survival in pancreatic ductal adenocarcinoma – a quantitative analysis. \textit{BMC Med. Imaging} 17, 38 (2017). 
\bibitem{cite9}Al-Kadi, O. \& Waston, D. Texture analysis of aggressive and nonaggressive lung tumor CE CT images. \textit{IEEE Trans. Biomed. Eng.} 55, 1822–1830 (2008). 
\bibitem{cite10}Haralick, R. M., Shanmugam, K. \& Dinstein, I. Textural features for image classification. \textit{IEEE Trans. Syst. Man Cybern.} 3, 610–621 (1973). 
\bibitem{cite11}Amadasun, M. \& King, R. Textural features corresponding to textural properties. \textit{IEEE Trans. Syst., Man, Cybern.} 19, 1264–1274 (1989). 
\bibitem{cite12}Galloway, M. Texture analysis using gray level run lengths. \textit{Computer Graph. Image Process.} 4, 172–179 (1975). 
\bibitem{cite13}Thibault, G. et al. Texture indexes and gray level size zone matrix. Application to cell nuclei classification. In\textit{10th International Conference On Pattern Recognition and Information Processing}. 140–145, (Minsk, Belarus, 2009). 

\bibitem{cite14}Lipkova, J. Artificial intelligence for multimodal data integration in oncology. \textit{Cancer Cell} 40, 1095–1110 (2022). 
\bibitem{cite15}Bashir, R. M. S., Qaiser, T., Raza, S. E. A. \& Rajpoot, N. M. HydraMix-Net: a deep multi-task semi-supervised learning approach for cell detection and classification. In \textit{Interpretable and Annotation-Efficient Learning for Medical Image Computing} (eds Cardoso, J. et al.) 164–171 (Springer International Publishing). \href{https://doi.org/10.1007/978-3-030-61166-8_18}{https://doi.org/10.1007/978-3-030-61166-8\_18} (2020). 
\bibitem{cite16}Gustavo, C., Iuri, A. \& Ricardo, C. A self-organizing map-based method for multi-label classification. \textit{IJCNN}, 4291–4298 (2017). 

\bibitem{cite17}Paul, R. Deep feature transfer learning in combination with traditional features predicts survival among patients with lung adenocarcinoma. \textit{Tomography} 2, 388–395 (2016). 

\bibitem{cite18}Liu, L. et al. Application of texture analysis based on apparent diffusion coefficient maps in discriminating different stages of rectal cancer. \textit{J. Magn. Reson Imaging} 45, 1798–1808 (2017). 

\bibitem{cite19}Vallières, M., Freeman, C. R., Skamene, S. R. \& El Naqa, I. A radiomics model from joint FDG-PET and MRI texture features for the prediction of lung metastases in soft-tissue sarcomas of the extremities. \textit{Phys. Med. Biol.} 60, 5471–5496 (2015). 

\bibitem{cite20}Doubeni, C. A. Precision screening for colorectal cancer: promise and challenges. \textit{Ann. Intern. Med.} 163, 390–391 (2015). 
\bibitem{cite21}Hao, Q. et al. Fusing multiple deep models for in vivo human brain hyperspectral image classification to identify glioblastoma tumor. \textit{IEEE Trans. Instrum. Meas}. 70, 4007314 (2021). 

\bibitem{cite22}Sun, W., Zheng, B. \& Qian, W. Automatic feature learning using multichannel ROI based on deep structured algorithms for computerized lung cancer diagnosis. \textit{Computers Biol. Med.} 89, 530–539 (2017). 

\bibitem{cite23}Huang, S.-C. Self-supervised learning for medical image classification: a systematic review and implementation guidelines. \textit{npj Digital Med.} 6, 74 (2023). 

\bibitem{cite24}Wang, C. et al. Non-invasive measurement using deep learning algorithm based on multi-source features fusion to predict PD-L1 expression and survival in NSCLC. \textit{Front. Immunol.} 13, 828560 (2022). 
\bibitem{cite25}Noorbakhsh, J. et al. Deep learning-based cross-classifications reveal conserved spatial behaviors within tumor histological images. \textit{Nat. Commun.} 11, 6367 (2020). 

\bibitem{cite26}Tomita, N. et al. Attention-Based Deep Neural Networks for Detection of Cancerous and Precancerous Esophagus Tissue on Histopathological Slides. \textit{JAMA Netw. Open} 2, e1914645 (2019). 

\bibitem{cite27}Xu, H. L. et al. Artificial intelligence performance in image-based ovarian cancer identification: A systematic review and meta-analysis. \textit{EClinicalMedicine} 53, 101662 (2022). 
\bibitem{cite28}Coudray, N. et al. Classification and mutation prediction from non–small cell lung cancer histopathology images using deep learning. \textit{Nat. Med.} 24, 1559–1567 (2018). 
\bibitem{cite29}Wang, S. et al. ConvPath: a software tool for lung adenocarcinoma digital pathological image analysis aided by a convolutional neural network. \textit{EBioMedicine} 50, 103–110 (2019). 

\bibitem{cite30}Girschik, J. et al. Precision in setting cancer prevention priorities: synthesis of data, literature, and expert opinion. \textit{Front. Public Health} 5, 125 (2017). 

\bibitem{cite31}Florimbi, G. et al. Accelerating the K-nearest neighbors filtering algorithm to optimize the real-time classification of human brain tumor in hyperspectral images. \textit{Sensors} 18, 2314 (2018). 

\bibitem{cite32}Wang, Y. et al. PARP inhibitors in gastric cancer: beacon of hope. \textit{J. Exp. Clin. Cancer Res}. 40, 211 (2021). 

\bibitem{cite33}He, K., Zhang, X., Ren, S. \& Sun, J. Deep residual learning for image recognition. In \textit{Proc. IEEE conference on computer vision and pattern recognition}, 770–778 (2016). 
\bibitem{cite34}Wang, X. et al. A prognostic and predictive computational pathology image signature for added benefit of adjuvant chemotherapy in early stage non-small-cell lung cancer. \textit{EBioMedicine} 69, 103481 (2021). 
\bibitem{cite35}Choi, S. et al. Deep learning model improves tumor-infiltrating lymphocyte evaluation and therapeutic response prediction in breast cancer. \textit{NPJ Breast Cancer} 9, 71 (2023). 

\bibitem{cite36}Otsu, N. A threshold selection method from gray-level histograms. \textit{IEEE Transactions on Systems, Man, and Cybernetics} 9, 62–66 (1979). 
\bibitem{cite37}Dietterich, T. G. Ensemble methods in machine learning. in \textit{Multiple Classifier Systems}. 1–15 (Springer Nature, 2000). 
\bibitem{cite38}Wang, H. et al. In \textit{Proc. IEEE/CVF Conference on Computer Vision and Pattern Recognition Workshops}. Preprint at \href{https://doi.org/10.48550/arXiv.1910.01279}{https://doi.org/10.48550/arXiv.1910.01279} (2019). 
\bibitem{cite39}He K., Zhang, X., Ren, S. \& Sun, J. Deep residual learning for image recognition. \textit{Proceedings of the IEEE Conference on Computer Vision and Pattern Recognition (CVPR)}. 770-778 (2016). 
\bibitem{cite40}Komura, D. \& Ishikawa, S. Machine learning methods for histopathological image analysis. \textit{Comput Struct. Biotechnol. J.} 16, 34–42 (2018). 

\bibitem{cite41}Camps-Valls, G. et al. Kernel-based methods for hyperspectral image classification. \textit{IEEE Trans. Geosci. Remote Sens.} 43, 1351–1362 (2005). 
Urbanos, G. et al. Supervised Machine \bibitem{cite42}Learning Methods and Hyperspectral Imaging Techniques Jointly Applied for Brain Cancer Classification. \textit{Sensors 2021} 21, 3827 (2021). 
\bibitem{cite43}Chen, L.-C., Papandreou, G., Shroff, F. \& Adam, H. Rethinking atrous convolution for semantic image segmentation. \textit{arXiv} \href{https://doi.org/10.48550/arXiv.1706.05587}{https://doi.org/10.48550/arXiv.1706.05587} (2017) 

\bibitem{cite44}Lu, M. Y., Sater, H. A. \& Mahmood, F. Multiplex computational pathology for treatment response predication. \textit{Cancer Cell} 39, 1053–1055 (2021). 
\end{thebibliography}
\end{document}